\documentstyle[psfig]{caosp}

\begin{document}
%k
%\pubyear{1998}
%\volume{27}
%\firstpage{458}
%
\hauthor{G. Scholz {\it et al.}}
\title{Magnetic field and element surface distribution of the CP2 star 
$\alpha^{2}$~CVn}
\author{Yu.V. Glagolevskij
\inst{1}
\and E. Gerth
\inst{2}
\and G. Hildebrandt
\inst{2}
\and H. Lehmann
\inst{3}
\and G. Scholz
\inst{2}}
\institute{Special Astrophysical Observatory of Russian AS, Nizhnij Arkhyz
357147, Russia\\
\and Astrophysikalisches Institut Potsdam, Telegrafenberg A27, D-14473
Potsdam, Germany\\
\and Th{\"u}ringer Landessternwarte Tautenburg,
Karl-Schwarzschild-Observatorium,
D-07778 Tautenburg, Germany}

%\date{December 12, 1997}

\maketitle

\begin{abstract}
We investigate the radial velocity and the magnetic field of the CP star 
$\alpha^{2}$~CVn. The observed variation of the magnetic field is compared with
that of our model. We search for a relation between the magnetic field 
and the distribution of the chemical elements. The period in the radial 
velocities is constant over a time interval of about 100 years.

\end{abstract}

\section{Introduction}
HD 112413 = $\alpha^{2}$~CVn~is a CP star whose variability  
in line intensities and radial velocities (RVs), was detected already 100 
years ago. In the earlier 50-ies $\alpha^{2}$~CVn was one of the first stars 
in which a magnetic field was discovered. Photographic observations of the 
longitudinal magnetic field $B_{\rm eff}$,~e.g. by Babcock \& Burd (1952), 
Oetken et al. (1970), gave a very anharmonic field curve whereas the 
photoelectric ones by Borra and Landstreet (1977) show a more harmonic shape.\\
Michaud (1970) and Glagolevskij (1994) showed that the distribution of 
the chemical elements on the stellar surface of magnetic stars has to be 
connected with the magnetic field. Therefore, we search for a relation 
between the concentration of the chemical elements and the structure of the 
magnetic field assuming a concrete magnetic field model.\\

\section{Observations}
For the investigation we use spectroscopic observations obtained with the 
\'echelle spectrograph and Zeeman analyzer (TRAFICOS, Hildebrandt et al. 1997) 
attached to the 2~m telescope of the Karl-Schwarzschild-Observatory at 
Tautenburg. Table 1 summarizes the results of the observations. The data are 
mean values of about 15 metallic lines.
\begin{table*}
\caption{\label{RV}RV and $B_{\rm eff}$ of $\alpha^{2}$~CVn}
\begin{flushleft}
\begin{tabular}{lcr|lcr}
\hline
HJD      &  RV             &  $B_{\rm eff}$ & HJD      &  RV            &
 $B_{\rm eff}$   \\
2450000+ & (km/s)          &  (gauss)       & 2450000+ & (km/s)         & 
 (gauss)        \\
\hline
494.658  & $-$0.3$\pm$0.6  &   +940$\pm$220 & 559.455  & $-$1.3$\pm$0.5 &
 $-$530$\pm$140  \\                 
496.646  & $-$3.3$\pm$0.6  & $-$720$\pm$280 & 583.378  & $-$1.6$\pm$0.4 &
   +690$\pm$280  \\          
528.513  & $-$2.4$\pm$0.8  &  +1070$\pm$330 & 585.347  & $-$6.7$\pm$0.9 &
 $-$960$\pm$220  \\
556.427  & $-$4.3$\pm$1.0  & $-$500$\pm$180 & 586.391  & $-$3.1$\pm$0.6 &
 $-$920$\pm$180  \\
558.420  & $-$4.5$\pm$0.9  & $-$990$\pm$220 & 588.364  & $-$2.0$\pm$0.6 &
  +1340$\pm$330  \\ 
\hline
\end{tabular}
\end{flushleft}
\end{table*}

\section{Radial velocities}
Radial velocities of $\alpha^{2}$~CVn~were accumulated over a period of 
nearly a century. For testing the validity of a unique ephemeris we 
investigate three sets of RVs. A frequency analysis gives no convincing 
evidence of a significant phase deviation between the different data sets. 
We find the same period of 5\fd469 given already by Farnsworth (1932), 
which is the period of rotation.

\section{Magnetic model of $\alpha^{2}$~CVn}
For the modelling we fit a calculated $B_{\rm eff}$-curve to the observed one.  
The calculated curve results from the superposition of a dipole with a higher 
multipole including a fixed combination of the angles $i$~and $\beta$; $i$~is
the inclination between the rotation axis 
and the line of sight and $\beta$ is the angle between the rotation axis 
and the magnetic axis. From the stellar parameters we estimate $i$~= 135\degr 
and we assume $\beta$~=~90\degr. The magnetic sources are located at 
$R_{\rm *}$~=~0.1 from the centre; they produce at the stellar surface the 
magnetic field strengths of the dipole of +3.2 and $-$3.2~kG and of the 
quadrupole of $-$6.0, +6.0, $-$6.0, and +6.0~kG. In Fig. 1 
the observed $B_{\rm eff}$-values (dots) and our model-curve (curve) 
are represented . 
 
\begin{figure}
%\picplace{5cm}
\psfig{figure=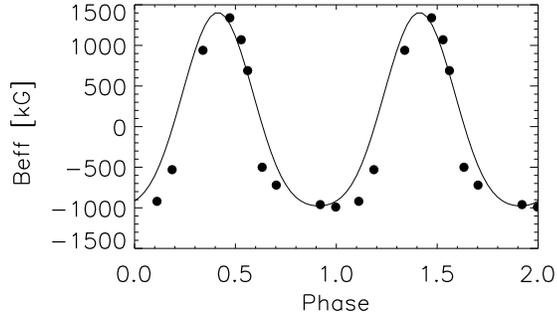,height=5cm}
\caption{\label{B}Variation of the longitudinal magnetic field. Measurements  
from Tautenburg (dots) and our model (curve)}
\end{figure}

\section{Magnetic structures and surface maps}
We take the Fe, Cr, and Ti distribution from the maps derived by Pyper (1969) 
and Khokhlova \& Pavlova (1984). These elements are mainly concentrated in 
the regions which coincide approximately with the longitudes of the negative 
quadrupoles. In other words, in our dipole-quadrupole model the positive and 
negative poles of the dipole are lying between the abundance spots, i.e., the 
elements are concentrated in a band around the zero magnetic field.

\section{Discussion}
Two findings seem to be important:\\ 
$-$~the rotation period, which is constant for nearly one hundred years,
suggests that the abundance patches have remained at a fixed position\\ 
$-$~the Fe, Cr, Ti distributions show quite a unique relation to the 
magnetic field.\\
   
\acknowledgements
This work was supported by the Th{\"u}ringer
Landessternwarte and its director Prof.~J.~Solf by allocation of observing 
time with our \'echelle Zeeman spectrograph. 

{}


\begin{thebibliography}{}
\bibitem{} Babcock H.W., Burd S., 1952, ApJ 116, 8
\bibitem{} Borra E.F., Landstreet J.D., 1977, ApJ 212, 141
\bibitem{} Farnsworth G., 1932, ApJ 75, 364
\bibitem{} Glagolevskij Yu.V., 1994, AZh 71, 858 
\bibitem{} Hildebrandt G., Scholz G., Rendtel J., Woche M., Lehmann H., 1997, 
Astron. Nachr. 318, 291
\bibitem{} Khokhlova V.L., Pavlova B.M., 1984, Pisma AZh 10, 377
\bibitem{} Michaud G., 1970, ApJ 160, 641
\bibitem{} Oetken L., Bartl E., Orwert R., 1970, Astron. Nachr. 292, 1
\bibitem{} Pyper D., 1969, ApJS 9, 321
\end{thebibliography}
\end{document}